\documentclass[reprint, longbibliography, superscriptaddress, secnumarabic, amssymb, nobibnotes, aps, prl, showkeys]{revtex4-1}

\usepackage{graphicx}
\usepackage{epstopdf}
\usepackage[T1]{fontenc}
\usepackage{amsbsy}
\usepackage{gensymb}
\usepackage[T1]{fontenc}
\usepackage{amsmath}
\usepackage{amssymb}
\usepackage{bbm}
\usepackage{braket}
\usepackage{xcolor}
\allowdisplaybreaks
\usepackage{graphicx}
\usepackage{hyperref}  
\hypersetup{
    unicode=false,          
    pdftoolbar=true,        
    pdfmenubar=true,        
    pdffitwindow=false,     
    pdfstartview={FitH},    
    pdftitle={Superconductivity in structurally complex $\sigma$-Phase Re-X (X = V, Nb, Ta) and a derived medium-entropy alloys},
    pdfauthor={},      
    pdfsubject={},   
    pdfcreator={},   
    pdfproducer={}, 
    pdfkeywords={} {} {}, 
    pdfnewwindow=true,      
    colorlinks=true,       
    linkcolor=blue, 
    citecolor=blue,        
    filecolor=magenta,      
    urlcolor=blue           
}

\usepackage[normalem]{ulem}
\usepackage{textcomp} 

\renewcommand{\approx}{\simeq}

\begin{document}
\title{Superconductivity in structurally complex $\sigma$-Phase Re-X (X = V, Nb, Ta) and a derived medium-entropy alloys}

\author{Pavan Kumar Meena}
\email{pavan.meena@pg.edu.pl}
\affiliation{Faculty of Applied Physics and Mathematics and Advanced Materials Centre, Gdansk University of Technology, Narutowicza 11/12, Gdansk 80-233, Poland}

\author{Tomasz Klimczuk}
\email{tomasz.klimczuk@pg.edu.pl}
\affiliation{Faculty of Applied Physics and Mathematics and Advanced Materials Centre, Gdansk University of Technology, Narutowicza 11/12, Gdansk 80-233, Poland}

\begin{abstract}
The structurally complex tetragonal sigma ($\sigma$)-phase provides a unique platform for investigating the interplay between chemical disorder, electronic structure, and superconductivity, particularly in Re-based alloys where unconventional superconductivity has been widely discussed. Here, we synthesize and investigate the superconducting properties of $\sigma$-phase Re-X (X = V, Nb, and Ta) alloys with compositions Re$_{0.76}$V$_{0.24}$, Re$_{0.56}$Nb$_{0.44}$, and Re$_{0.60}$Ta$_{0.40}$. These compositions lie within the narrow stability range of the $\sigma$ phase, underscoring the critical role of valence electron concentration in phase formation. To examine the influence of enhanced chemical disorder, we further synthesize the structurally complex medium-entropy alloy Re$_{0.56}$Nb$_{0.19}$Ta$_{0.19}$V$_{0.06}$, derived from these binary systems. Magnetization, electrical resistivity, and specific-heat measurements establish bulk type-II superconductivity in all compounds. Analysis of the electronic heat capacity is consistent with a fully gapped, weak-coupling BCS superconducting state. In contrast, the normal-state resistivity exhibits an unconventional negative temperature coefficient, and the superconducting transition temperature values obtained from resistivity measurements are higher than those determined from other measurements. Our results demonstrate that both the $\sigma$-phase Re-X alloys, spanning 3d, 4d, and 5d transition-metal substitutions, and their medium-entropy counterpart constitute an attractive family of model systems for investigating the effects of structural complexity, chemical disorder, and spin-orbit coupling on superconductivity in Re-based materials.
\end{abstract}
\maketitle

\section{INTRODUCTION}

The sigma ($\sigma$) phase is a structurally complex tetragonal intermetallic characterized by five crystallographically nonequivalent atomic sites that accommodate substantial chemical substitution and compositional disorder \cite{hall1966sigma, joubert2008crystal, shoemaker1950crystal, bergman1954determination, yaqoob2012comparison, kasper1956ordering}. Owing to its structural complexity and sensitivity to valence electron concentration (VEC), the $\sigma$ phase provides an attractive platform for investigating the interplay between crystal structure, electronic structure, and emergent physical properties \cite{koch1971superconductivity,khan1980superconductivity}. Although generally brittle, $\sigma$-phase compounds exhibit high mechanical hardness and remarkable compositional flexibility, making them relevant for both functional and structural materials research \cite{singhal1968formation, hsieh2012overview, tavares2010magnetic}.

Superconductivity has been reported in numerous transition-metal $\sigma$-phase compounds \cite{koch1971superconductivity, khan1980superconductivity, shang2019structure, shang2020re1, carnicom2017new, carnicom2018sigma}, where the superconducting transition temperature ($T_c$) is strongly correlated with the electronic density of states and valence electron concentration \cite{blaugher1961superconductivity, khan1980superconductivity, joubert2008crystal, roberts1967intermetallic, blaugher1961superconductivity, rasmussen1987new, khan1978comparison, compton1961superconductivity, khan1977comparison, khan1979superconductivity}. However, the narrow compositional stability of many $\sigma$-phase compounds has hindered systematic investigations of how chemical substitution, electronic structure, and lattice disorder collectively influence superconductivity. Consequently, material systems that retain the same $\sigma$-phase framework while allowing controlled variations in composition remain scarce. Among these materials, Re-based superconductors are of particular interest because several members of this family have been reported to exhibit unconventional superconductivity and time-reversal symmetry breaking, although the microscopic origin of these phenomena remains unresolved \cite{ghosh2021recent, shang2018time, kushwaha2024unconventional, meena2026superconducting, shang2019structure, shang2020re1}. Proposed explanations include strong spin-orbit coupling, structural complexity, chemical disorder, and the intrinsic electronic character of Re. Despite this interest, superconductivity in Re-based $\sigma$-phase compounds has received comparatively little attention, and systematic studies across different transition-metal substitutions remain scarce.

\begin{figure*}
\includegraphics[width=2.0\columnwidth]{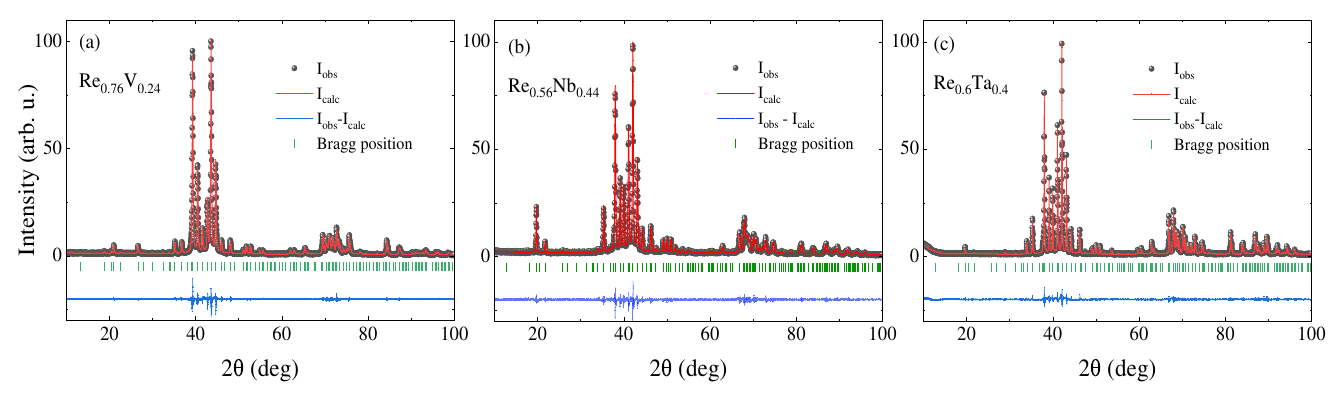}
\caption {\label{Fig1} Powder X-ray diffraction patterns of (a) Re$_{0.76}$V$_{0.24}$, (b) Re$_{0.56}$Nb$_{0.44}$, and (c) Re$_{0.60}$Ta$_{0.40}$. Experimental data are shown as symbols, while the solid red lines represent the Le Bail refinements, confirming the tetragonal $\sigma$-phase structure of all compounds. The difference between observed and calculated intensities is shown below of each patterns. Vertical tick marks indicate the expected Bragg reflection positions corresponding to the allowed $(hkl)$ planes.}
\end{figure*}

In this work, we synthesize and systematically investigate the superconducting properties of the structurally complex $\sigma$-phase alloys Re$_{0.76}$V$_{0.24}$, Re$_{0.56}$Nb$_{0.44}$ and Re$_{0.60}$Ta$_{0.40}$, representing transition-metal substitutions of 3d, 4d and 5d, respectively, thus allowing a systematic variation of spin-orbit coupling while preserving the same crystal structure. The $\sigma$ phase is known to occupy a narrow stability window in the phase diagram, making its synthesis highly sensitive to composition and requiring precise control of valence electron concentration. Although superconductivity has been reported in several Re-X $\sigma$-phase alloys since the earliest studies, these reports are often limited by partial phase characterization and incomplete compositional information. Systematic investigations of their superconducting state, together with precise compositional ranges required to stabilize the $\sigma$ phase, remain scarce \cite{jorda1986vanadium, bucher1961supraleitung, knapton1959niobium, joubert2008crystal, roberts1976survey}. To further explore the influence of chemical disorder without altering the underlying crystal structure, we also synthesized a structurally complex medium-entropy alloy (MEA) Re$_{0.56}$Nb$_{0.19}$Ta$_{0.19}$V$_{0.06}$ derived from the parent binary compounds. Magnetization, electrical resistivity, and specific-heat measurements confirm bulk type-II superconductivity with transition temperatures T$_c$ $\approx$ 4.46, 2.28, 1.76 and 1.64 K for Re$_{0.76}$V$_{0.24}$, Re$_{0.56}$Nb$_{0.44}$, Re$_{0.60}$Ta$_{0.40}$ and Re$_{0.56}$Nb$_{0.19}$Ta$_{0.19}$V$_{0.06}$, respectively. The superconducting state is consistent with weak-coupling, fully gapped BCS behavior, while the normal state exhibits an unusual metallic response with a negative temperature coefficient and the $T_c$ values obtained from resistivity measurements are higher than those obtained from bulk measurements. Together, these results establish the Re-based $\sigma$-phase family and its medium-entropy derivative as a versatile model system for investigating how structural complexity, chemical disorder, spin-orbit coupling, and electronic structure influence superconductivity.

\begin{figure*}
\includegraphics[width=2.0\columnwidth]{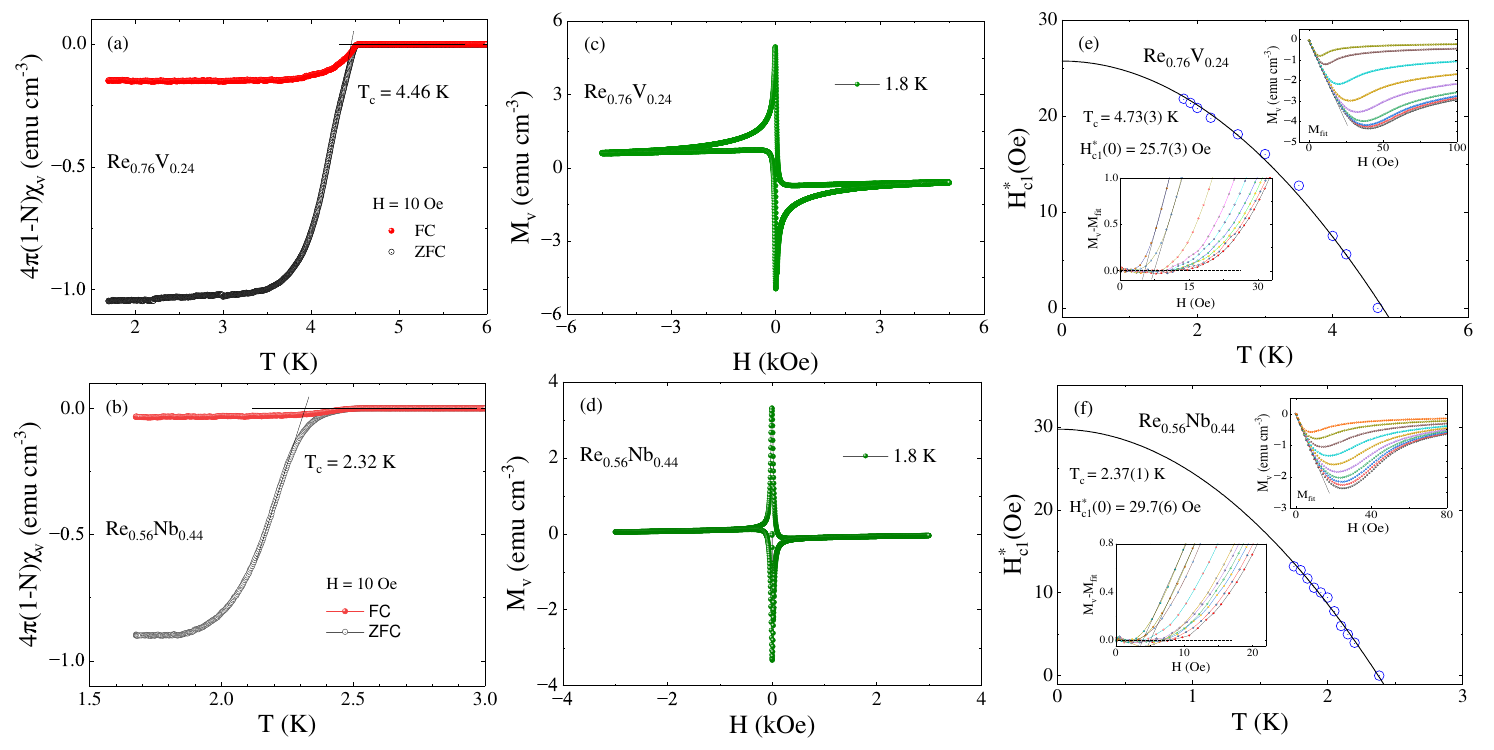}
\caption {\label{Fig2} (a) and (b) Zero-field-cooled (ZFC) and field-cooled (FC) volume magnetic susceptibility of Re$_{0.76}$V$_{0.24}$ and Re$_{0.56}$Nb$_{0.44}$ compounds measured in a magnetic field of 10 Oe, the transition from the measurements are 4.48(2) and 2.31(5) K for Re$_{0.76}$V$_{0.24}$ and Re$_{0.56}$Nb$_{0.44}$ compounds, respectively. (c) and (d) The magnetic field-dependent magnetization at 1.8 K is shown, confirming type-II superconductivity. (e) and (f) The estimation of $H^{*}_{c1}$ from the ($M_{V}$) vs applied field (H) isotherms as shown in the inset of (e) and (f). Temperature-dependent $H^{*}_{c1}$ fitted with Ginzburg-Landau relation. The critical temperature values at zero field were used from the heat capacity transition temperature.}
\end{figure*}

\section{EXPERIENTIAL DETAILS}
Polycrystalline samples of Re-X (X = V, Nb, and Ta) and a medium-entropy alloy Re-Nb-Ta-V were synthesized by arc melting under a high-purity argon atmosphere. A zirconium getter was used to further reduce residual oxygen during synthesis. Stoichiometric amounts of high-purity elements were melted on a water-cooled copper hearth and flipped and remelted 4-5 times to ensure chemical homogeneity, with minimal mass loss observed. Phase purity and crystal structure were examined using powder X-ray diffraction (pXRD) performed on a Bruker D2 Phaser diffractometer equipped with an XE-T detector and CuK$_\alpha$ radiation. Structural refinements were carried out using the Le Bail method within the Bruker TOPAS software package. The stoichiometric compositions were confirmed by energy-dispersive X-ray spectroscopy (EDS) on the same samples used for other measurements, with three randomly selected points analyzed on a polished flat surface of each sample. Temperature-dependent magnetization, electrical resistivity, and heat capacity measurements were performed using a Quantum Design Dynacool Physical Property Measurement System (PPMS). Heat capacity and resistivity measurements were conducted using a $^3$He insert to access low temperatures, while magnetic measurements were performed using the $^4$He option. Heat capacity was measured via the standard thermal relaxation method, and electrical resistivity was determined using a conventional four-probe technique with platinum wires spot-welded to polished sample surfaces to ensure reliable electrical contacts.

\section{RESULTS}

\subsection{Binary alloys}
Room-temperature powder X-ray diffraction (pXRD) measurements were performed on Re-X (X = V, Nb, and Ta) compounds to investigate their crystal structures and phase purity. The diffraction patterns were analyzed using the Le Bail refinement method, as the crystal structure of these binary $\sigma$-phase compounds is already well established; the refinement was used to confirm the phase and determine the lattice parameters, and the results are shown in the Fig.~\ref{Fig1}(a-c). All three compounds, with compositions Re$_{0.76}$V$_{0.24}$, Re$_{0.56}$Nb$_{0.44}$, and Re$_{0.60}$Ta$_{0.40}$, were found to crystallize in a tetragonal CrFe-type structure. The corresponding lattice parameters are summarized in Table~\ref{tbl:parameters}, which are in good agreement with reported in literature. The synthesis of these compounds is particularly challenging due to the narrow stability window of the tetragonal phase, where small deviations from the target composition can lead to the formation of secondary phases such as $\alpha$-Mn or W-type cubic structures. This strong compositional sensitivity highlights the crucial role of precise stoichiometry in stabilizing the $\sigma$ phase in Re-X systems. 
Energy-dispersive X-ray spectroscopy (EDS) confirms compositions of V = $26.0 \pm 1.6$ at.\% and Re = $74.0 \pm 1.6$ at. \% for Re$_{0.76}$V$_{0.24}$, Nb = $42.7 \pm 1.1$ at. \% and Re = $57.3 \pm 1.1$ at. \% for Re$_{0.56}$Nb$_{0.44}$, and Ta = $38.2 \pm 1.3$ at. \% and Re = $61.8 \pm 1.3$ at. \% for Re$_{0.60}$Ta$_{0.40}$. The uncertainties represent standard deviations across different sample locations, and the measured compositions closely match the nominal values. Scanning electron microscopy (SEM) reveals a uniform microstructure.

\begin{figure*}
\includegraphics[width=2.0\columnwidth]{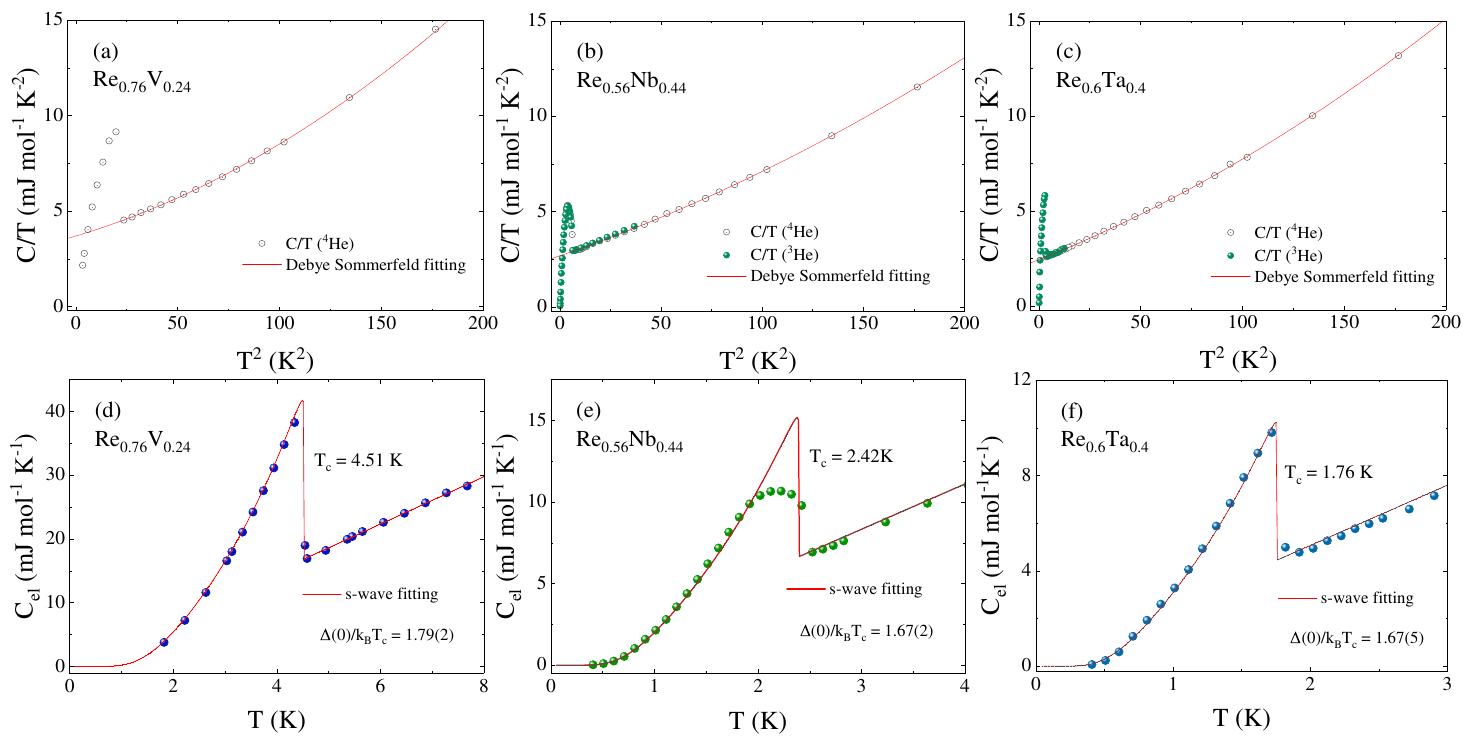}
\caption {\label{Fig3} (a-c) Temperature dependence of the zero-field heat capacity of  Re$_{0.76}$V$_{0.24}$, Re$_{0.56}$Nb$_{0.44}$ and Re$_{0.60}$Ta$_{0.40}$, presented as C/T versus T$^{2}$. A clear superconducting transition from the normal state is observed at the respective superconducting transition temperatures. (d-f) Temperature dependence of the electronic heat capacity C$_{el}$ as a function of T, extracted from the total heat capacity and fitted using an isotropic s-wave superconducting gap model.}
\end{figure*}

Temperature-dependent magnetization measurements were performed on Re$_{0.76}$V$_{0.24}$ and Re$_{0.56}$Nb$_{0.44}$ under an applied magnetic field of 1.0 mT using both zero-field-cooled (ZFC) and field-cooled (FC) protocols. The corresponding data are presented in Fig.~\ref{Fig2}(a) and (b). Both compounds exhibit a pronounced diamagnetic response below the superconducting transition, confirming the superconductivity. The superconducting transition temperatures, $T_c$, were determined from the intersection of the steepest slope of the ZFC susceptibility curve with the extrapolated normal-state background \cite{klimczuk2004carbon}, yielding $T_c$ = 4.49 K for Re$_{0.76}$V$_{0.24}$ and $T_c$ = 2.38 K for Re$_{0.56}$Nb$_{0.44}$. Magnetization data for Re$_{0.60}$Ta$_{0.40}$ are not available because its lower $T_c$ requires measurements in a $^3$He magnetometer. A clear bifurcation between the ZFC and FC curves is observed below $T_c$ in both compounds, which is attributed to the polycrystalline nature of the samples.

Evidence for type-II superconductivity is provided by field-dependent magnetization measurements, as shown in Fig.~\ref{Fig2}(c) and (d). The susceptibility data in Fig.~\ref{Fig2}(a) and (b) were corrected for demagnetization effects using the relation $-4\pi\chi_V = 1/(1-N)$, where the demagnetization factors are $N = 0.49$ and $0.46$ for Re$_{0.76}$V$_{0.24}$ and Re$_{0.56}$Nb$_{0.44}$, respectively. These values were estimated from low-temperature isothermal $M(H)$ measurements [insets of Fig.~\ref{Fig2}(e) and (f)] and are consistent with those expected from the sample geometry. Following correction, the ZFC susceptibility $\chi_V(T)$ approaches $-1$ at low temperatures, indicating complete magnetic shielding. In contrast, the considerably smaller FC signal reflects a reduced Meissner fraction, commonly observed in polycrystalline superconductors owing to strong flux pinning at grain boundaries.

To determine the lower critical field, $H_{c1}(0)$, a series of low-field isothermal magnetization ($M(H)$) measurements were performed at various temperatures within the superconducting state. In the Meissner region, each $M(H)$ curve [insets of Fig.~\ref{Fig2}(e) and (f)] was analyzed to identify the field at which the magnetization deviates from the initial linear response. Since the onset cannot be reliably identified visually, we quantified it from $M-M_{\mathrm{fit}}$, where $M_{\mathrm{fit}}$ is the low-field linear fit. The field at which $M-M_{\mathrm{fit}}$ exhibits a systematic deviation from zero was taken as the onset of flux penetration and defined as the effective lower critical field, $H_{c1}^{*}(T)$, as shown in the lower insets of Fig.~\ref{Fig2}(e) and (f). This procedure follows the method described in Ref.~\cite{xu2022superconductivity}. This deviation was taken as the effective lower critical field, $H_{c1}^{*}(T)$. The resulting temperature dependence of $H_{c1}^{*}(T)$ is shown in the main panels of Fig.~\ref{Fig2}(e) and (f). The data were analyzed using the expression $H^{*}_{c1}(T) = H^{*}_{c1}(0)(1 - (T/T_{c})^{2}$), which yielded $H_{c1}^{*}(0)$ = 25.7(3) Oe and 32.0(2) Oe for Re$_{0.76}$V$_{0.24}$ and Re$_{0.56}$Nb$_{0.44}$, respectively \cite{poole1995superconductivity}. After correcting for demagnetization effect according to $H_{c1}(0)=H_{c1}^{*}(0)/(1-N)$, the intrinsic lower critical fields were determined to be $H_{c1}(0) = 51.0(2)$ Oe for Re$_{0.76}$V$_{0.24}$ and $60.1(5)$ Oe for Re$_{0.56}$Nb$_{0.44}$. These values were subsequently employed in the determination of the remaining superconducting parameters.

The zero-field low-temperature heat-capacity measurements are shown in Fig. \ref{Fig3}(a-f) and are plotted as $C/T$ versus $T^2$ in Fig. \ref{Fig3}(a-c). In the normal state, the data were analyzed using the Debye-Sommerfeld expression C/T = $\gamma_{n}$ +$\beta_{3}$ T$^2$ + $\beta_{5} T^{4}$, where $\gamma_n T$ represents the electronic contribution, $\beta_3 T^3$ arises from lattice vibrations, and $\beta_5 T^5$ accounts for higher-order anharmonic lattice contributions. The solid red lines represent fits to the experimental data, and the extracted parameters are summarized in Table \ref{tbl:parameters}. 

Within the Debye model, the coefficient $\beta_{3}$ is related to the Debye temperature $\Theta_{D}$ through
\begin{equation}
\Theta_{D}=\left(\frac{12\pi^{4}nR}{5\beta_{3}}\right)^{1/3},
\label{Debye}
\end{equation}
where $n$ is the number of atoms per formula unit ($n = 1$ for Re$_{0.76}$V$_{0.24}$, Re$_{0.56}$Nb$_{0.44}$, and Re$_{0.60}$Ta$_{0.40}$) and $R = 8.314$ J mol$^{-1}$ K$^{-1}$ is the molar gas constant. Using Eq. (\ref{Debye}), the Debye temperatures were estimated to be 394, 377 and 358 K for Re$_{0.76}$V$_{0.24}$, Re$_{0.56}$Nb$_{0.44}$, and Re$_{0.60}$Ta$_{0.40}$, respectively. The Debye temperature decreases systematically with the substitution of V by Nb and Ta, consistent with the corresponding change in atomic mass and in agreement with theoretical expectations.

\begin{figure*}
\includegraphics[width=2.0\columnwidth]{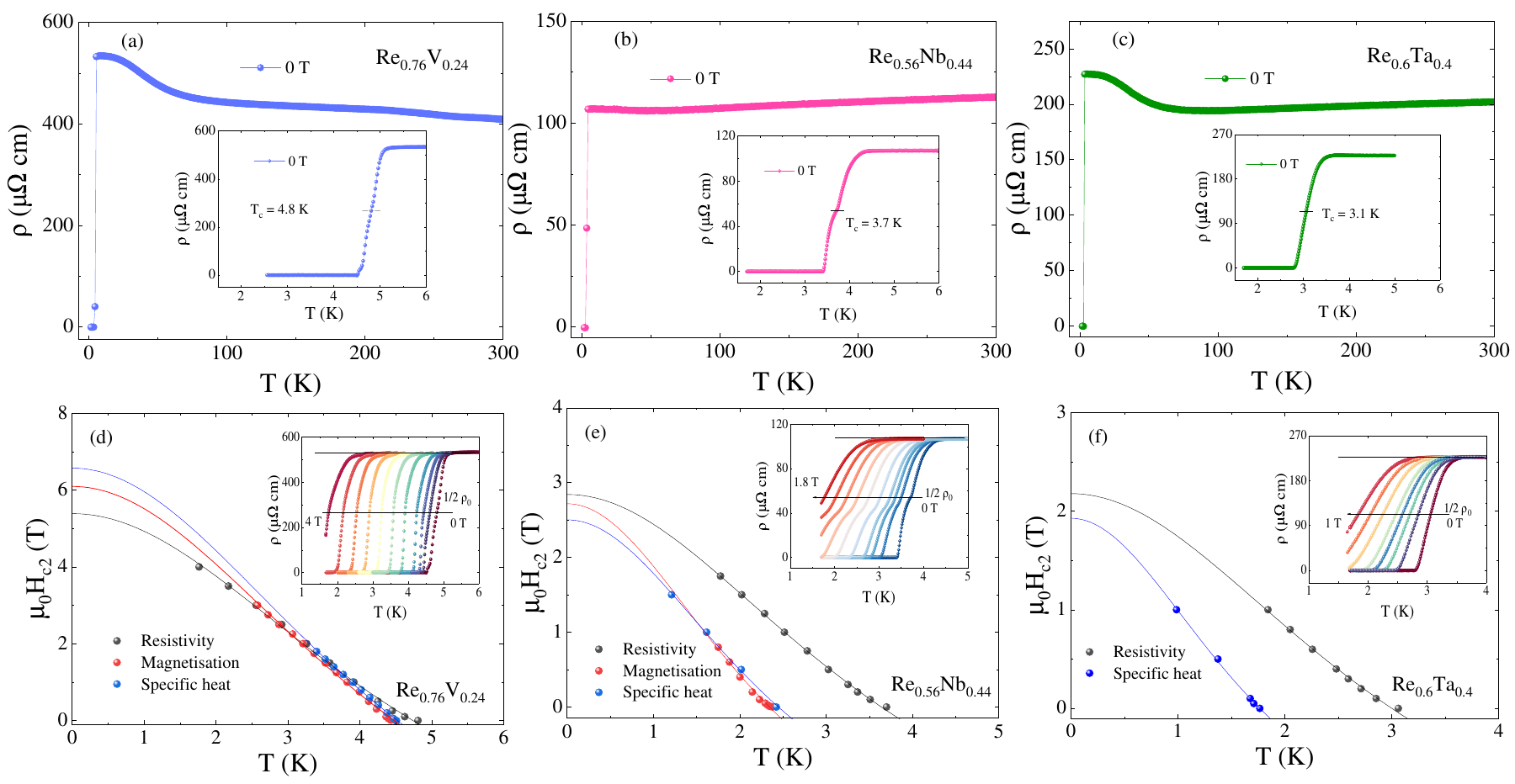}
\caption {\label{Fig4} (a-c) Temperature dependence of resistivity for Re$_{0.76}$V$_{0.24}$, Re$_{0.56}$Nb$_{0.44}$, and Re$_{0.60}$Ta$_{0.40}$. The inset shows the zoomed view of the resistivity, showing a clear transition in resistivity. (d-f) The upper critical field estimated from resistivity, magnetization, and heat capacity measurements for these compounds is shown, fitted with the Gingzburg Landau relation, where the inset shows temperature-dependent resistivity data with applied different magnetic fields.}
\end{figure*}

Using the estimated Debye temperatures $\Theta_D$ and assuming a Coulomb pseudopotential $\mu^{*}=0.13$, a value commonly adopted for intermetallic superconductors, the electron-phonon coupling constant $\lambda_{\mathrm{el-ph}}$ was evaluated using the inverted McMillan equation \cite{mcmillan1968transition},
\begin{equation}
\lambda_{\mathrm{el-ph}} =
\frac{1.04 + \mu^{*}\ln \left(\Theta_D/1.45T_c\right)}
{(1-0.62\mu^{*})\ln \left(\Theta_D/1.45T_c\right)-1.04}.
\label{eqn6}
\end{equation}
The resulting values of $\lambda_{\mathrm{el-ph}}$ are 0.57(4), 0.50(2), and 0.48(2) for Re$_{0.76}$V$_{0.24}$, Re$_{0.56}$Nb$_{0.44}$, and Re$_{0.60}$Ta$_{0.40}$, respectively. These values are characteristic of superconductors with weak electron-phonon coupling. The systematic decrease in $\lambda_{\mathrm{el-ph}}$ across the series correlates with the reduction in $T_c$, suggesting a gradual weakening of the superconducting pairing interaction from Re$_{0.76}$V$_{0.24}$ to Re$_{0.60}$Ta$_{0.40}$.

Having both $\gamma_n$ and $\lambda_{\mathrm{el-ph}}$ we can estimate the electronic density of states by using the following expression: 
\begin{equation}
DOS(E_{F}) = \frac{3\gamma_{n}}{\pi^{2} k_{B}^{2}(1+\lambda_{el-ph})},
\label{DOS}
\end{equation}
where $k_B = 1.38 \times 10^{-23}$ J K$^{-1}$ is the Boltzmann constant. The resulting values of $DOS(E_F)$ are 1.0, 0.76, and 0.70 states eV$^{-1}$ f.u.$^{-1}$ for Re$_{0.76}$V$_{0.24}$, Re$_{0.56}$Nb$_{0.44}$, and Re$_{0.60}$Ta$_{0.40}$, respectively.

The electronic contribution to the heat capacity, $C_{\mathrm{el}}$, was obtained by subtracting the lattice contribution from the total heat capacity measured in zero magnetic field, through Debye-Sommerfeld expression, yielding $C_{\mathrm{el}} = C - \beta_{3}T^{3} - \beta_{5}T^{5}$. Temperature dependent $C_{\mathrm{el}}$ for the binary Re$_{0.76}$V$_{0.24}$, Re$_{0.56}$Nb$_{0.44}$, and Re$_{0.60}$Ta$_{0.40}$ is shown for in Fig. \ref{Fig3}(d-f). All three compounds exhibit a pronounced anomaly at the superconducting transition, providing clear evidence for the bulk nature of superconductivity. The superconducting transition temperature, $T_c$, was determined using the equal-entropy construction, which ensures entropy balance between the normal and superconducting states across the transition. The resulting values of $T_c$ are 4.46, 2.28 and 1.76 K for Re$_{0.76}$V$_{0.24}$, Re$_{0.56}$Nb$_{0.44}$, and Re$_{0.60}$Ta$_{0.40}$, respectively, in good agreement with those obtained from magnetization measurements and reported in literature \cite{roberts1976survey}. Having the obtained Sommerfeld coefficient $\gamma_n$, the normalized specific-heat jump at $T_c$, $\Delta C/\gamma_n T_c$, was estimated to be 1.52(1), 1.53(1), and 1.54(2) for Re$_{0.76}$V$_{0.24}$, Re$_{0.56}$Nb$_{0.44}$, and Re$_{0.60}$Ta$_{0.40}$, respectively. These values are slightly larger than the weak-coupling BCS prediction of 1.43 \cite{tinkham2004introduction}, suggesting weak to moderate coupling superconductivity.

\begin{table*} 
\caption{Refined unit cell parameters and Normal and superconducting parameters of the binary alloys Re$_{0.76}$V$_{0.24}$, Re$_{0.56}$Nb$_{0.44}$ and Re$_{0.60}$Ta$_{0.40}$ and the derived medium entropy alloy Re$_{0.56}$Nb$_{0.19}$Ta$_{0.19}$V$_{0.06}$ obtained from x-ray refinement, magnetization, resistivity, and heat capacity measurements. }
\label{tbl:parameters}
\setlength{\tabcolsep}{13pt}
\begin{center}
\begin{tabular}[b]{l c c c c c c c}
\hline
\hline
Parameters& unit& Re$_{0.76}$V$_{0.24}$ & Re$_{0.56}$Nb$_{0.44}$ & Re$_{0.60}$Ta$_{0.40}$ &Re$_{0.56}$Nb$_{0.19}$Ta$_{0.19}$V$_{0.06}$\\
\hline
Structure & \multicolumn{5}{c}{Tetragonal $\beta$-CrFe} \\ 
Space group  & \multicolumn{5}{c}{P4$_2$/mnm (No. 136)} \\
a = b &\AA & 9.4656(1) & 9.7723(1) & 9.7550(1) &9.7309(3)\\
c &\AA & 4.8916(2) & 5.0878(5) &5.0820(6) &5.0650(7)\\
$V_{cell}$ &\AA$^{3}$ & 438.28(1) &  485.88(1) & 483.61(1) &479.60(6)\\
$T_{c}$ & K & 4.46 & 2.28 &1.76 &1.64\\
$\mu_{0}H_{c1}(0)$ & mT & 5.1(2) & 6.0(5) &  &\\ 
$\mu_{0}H_{c2}$(0) & T & 6.57(7) & 2.50(5) & 1.92(3) &2.24(3)\\
$\mu_{0}H_{c2}^{P}$(0) & T & 8.4 & 4.5 &3.3 &3.1\\
$\mu_{0}H_{c}$ & mT  & 92 & 68 & &\\
$\xi_{GL}$& \text{\AA} & 71 & 114 & 131 &   121 \\
$\lambda_{GL}^{mag}$& \text{\AA} & 3550 & 2980 & &\\
$k_{GL}$& & 50& 26& &\\
$\gamma_{n}$& mJ mol$^{-1}$ K$^{-2}$ & 3.70(3) & 2.68(1) & 2.44(1) & 2.59(7)\\
$\beta_{3}$& mJ mol$^{-1}$ K$^{-4}$ & 0.031(7) & 0.036(3) & 0.042(4) & 0.040(2)\\
$\beta_{5}$& $\mu$J mol$^{-1}$ K$^{-6}$ & 0.16(3) & 0.077(1) & 0.10(1) & 0.087(4)\\
$\theta_{D}$& K& 394 & 377 & 358 &365\\
$\frac{\Delta(0)}{k_{B}T_{c}}$ &  & 1.79(2) & 1.67(2) & 1.67(5) &1.45(4)\\
$\lambda_{e-ph}$ &  & 0.57 & 0.50 & 0.48 & 0.47\\
$\frac{\Delta C}{\gamma_{n} T_{c}}$ &  & 1.52(1) & 1.53(1) & 1.54(2) &1.05(2)\\
$DOS(E_{F})$ & st. eV$^{-1}$ f.u.$^{-1}$  & 1.00 & 0.76& 0.70 & 0.75\\
\hline
\hline
\end{tabular}
\par\medskip\footnotesize
\end{center}
\end{table*}

To analyze the superconducting state, the low-temperature electronic heat capacity data were examined within the framework of the $\alpha$-model based on BCS theory \cite{padamsee1973quasiparticle}. In this approach, the thermodynamic entropy $S$ is first calculated, and the electronic heat capacity is obtained from $C_{\mathrm{el}} = T dS/dT$. The normalized entropy in the superconducting state is given by
\begin{equation}
\frac{S}{\gamma_{n}T_{c}} = -\frac{6}{\pi^{2}} \left(\frac{\Delta(0)}{k_{B}T_{c}}\right)
\int_{0}^{\infty} \left[f\ln f + (1-f)\ln(1-f)\right] dy,
\label{BCS}
\end{equation}
where $f(\xi)$ = $(\exp(E(\xi)/k_{B}T)+1)^{-1}$ is the Fermi-Dirac distribution function, and $y = \xi/\Delta(0)$ is the integration variable. The quasiparticle excitation energy is given by $E(\xi) = \sqrt{\xi^{2} + \Delta^{2}(t)}$, where $\Delta(t)$ is the temperature-dependent superconducting gap and $t = T/T_{c}$ is the reduced temperature.

Within the weak-coupling BCS approximation, the temperature dependence of the superconducting gap is described by the empirical relation
\begin{equation}
\Delta(t)=tanh[1.82((1.018(1/t))-1)^{0.51}].
\end{equation}
The electronic heat capacity (C$_{el}$) data were described by the isotropic s-wave superconducting gap model (Fig. \ref{Fig3}(d-f)). The extracted zero-temperature superconducting gap ratios, $\alpha = \Delta(0)/k_{B}T_{c}$, was obtained as 1.79(2), 1.67(2), and 1.67(5) for Re$_{0.76}$V$_{0.24}$, Re$_{0.56}$Nb$_{0.44}$, and Re$_{0.60}$Ta$_{0.40}$, respectively. These values are close to the weak-coupling BCS limit of 1.74 \cite{bardeen1957theory}, indicating that all three compounds are consistent with weak-coupling superconductivity. While these results support conventional superconducting behavior, a definitive determination of the gap symmetry would require additional microscopic probes such as $\mu$SR. It is also noteworthy that Re-based superconductors are known to host unconventional superconducting states in some cases, despite exhibiting otherwise conventional bulk properties \cite{singh2014detection, ghosh2021recent, shang2018time, kushwaha2024unconventional, shang2019structure, shang2020re1, mandal2025time}.

\begin{figure*}
\includegraphics[width=2.0\columnwidth]{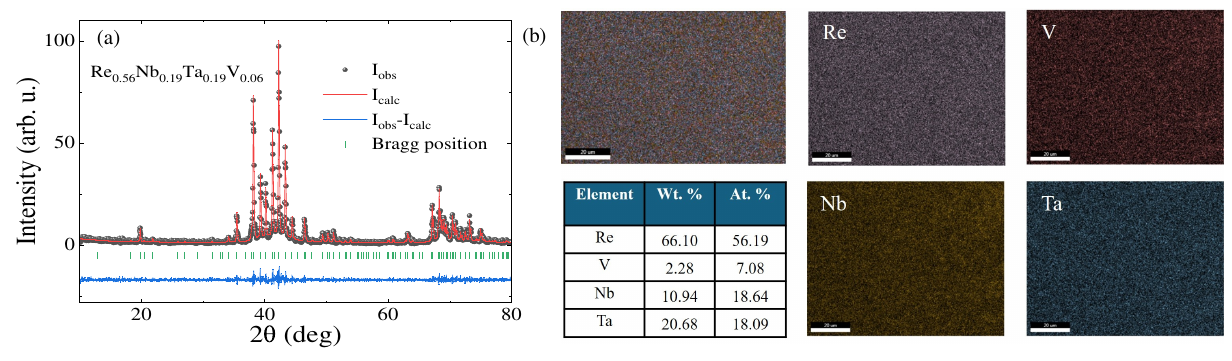}
\caption {\label{Fig5} (a) Powder XRD patterns of Re$_{0.56}$Nb$_{0.19}$Ta$_{0.19}$V$_{0.06}$, where experimental data and Le Bail fit shows by points and solid red line respectively, confirming the tetragonal structure of the compounds. The difference between the observed and calculated intensities is displayed beneath each pattern. Vertical tick marks indicate the expected Bragg peak positions corresponding to the respective hkl planes. (b) Elemental map of Re$_{0.56}$Nb$_{0.19}$Ta$_{0.19}$V$_{0.06}$ obtained using EDS, where different elements are represented by distinct colors, confirming the uniform distribution within the sample}
\end{figure*}

The temperature dependence of the electrical resistivity for all investigated samples is shown in Fig. \ref{Fig4}(a-c). At high temperature the Nb- and Ta-containing compounds exhibit only a weak metallic temperature dependence of the resistivity $\rho(T)$ with a minimum occurring at around 50 K. In contrast, the resistivity of the Re$_{0.76}$V$_{0.24}$ sample increases with decreasing temperature over the entire measured temperature range, although the increase becomes more pronounced below 50 K. A comparison of Fig. \ref{Fig4}(a-c) shows that the low-temperature upturn is strongest in Re$_{0.76}$V$_{0.24}$, suppressed in Re$_{0.56}$Nb$_{0.44}$, and enhanced again in Re$_{0.60}$Ta$_{0.40}$. This variation may reflect differences in chemical composition, Re content, and chemical/structural disorder. Similar behavior has been reported in several related intermetallic systems \cite{hulm1972superconductivity, hake1961electrical, chandrasekhar1961paramagnetic, khan1980superconductivity, carnicom2017superconductivity, carnicom2018sigma, fisk1973normal}, and is typically attributed to dominant impurity or vacancy scattering, consistent with Cohen’s model \cite{cohen1967effect}. Upon cooling, the resistivity of each compound drops to zero, confirming superconductivity. The transitions to the superconducting state are relatively broad in all compounds, which may reflect intrinsic disorder associated with the $\sigma$-phase structure. The superconducting transition temperatures $T_c$, defined as the midpoint of the transition and shown in the insets of Fig. \ref{Fig4}(a-c), are 4.81, 3.70 and 3.06 K for Re$_{0.76}$V$_{0.24}$, Re$_{0.56}$Nb$_{0.44}$, and Re$_{0.60}$Ta$_{0.40}$, respectively. While the critical temperature of Re$_{0.76}$V$_{0.24}$ agrees well with the $T_c$ values determined by other experimental methods, the transition temperatures obtained for the other two compounds are significantly higher. 

This difference may be associated from surface superconductivity \cite{han2013superconductivity, tsindlekht2004tunneling, kushwaha2026microscopic}, which can persist to somewhat higher temperatures than bulk superconductivity. However, filamentary superconductivity, disorder, or spatially inhomogeneous superconductivity may also contribute to the observed resistive response, and the precise origin cannot be established from the present measurements alone. In addition, Re$_{0.56}$Nb$_{0.44}$ exhibits a two-step superconducting transition in the resistivity over the entire measured magnetic-field range. Its persistence at high fields argues against impurity phases and instead suggests an origin related to inhomogeneous superconducting regions, although the absence of additional reflections in the pXRD pattern does not support the presence of secondary phases.

To determine the zero-temperature upper critical field $H_{c2}(0)$, magnetization, heat capacity and resistivity measurements were carried out under various applied magnetic fields. As expected, increasing the magnetic field progressively suppresses superconductivity, shifting the transition to lower temperatures in all three compounds, as shown in the insets of Fig. \ref{Fig4}(d-f). The resulting $H_{c2}(T)$ data are presented as a function of temperature in Fig. \ref{Fig4}(d)-(f). Within the phenomenological Ginzburg–Landau (GL) interpolation framework, the temperature dependence of the upper critical field is described by:

\begin{equation}
H_{c2}(T) = H_{c2}(0)\left[\frac{(1-t^{2})}{(1+t^{2})}\right];  \quad  \text{where} \;  t = \frac{T}{T_{\rm c}}.
\label{eqn3:HC1}
\end{equation}
where $t$ is the reduced temperature and $T_{c}$ is the zero-field superconducting transition temperature obtained from fitting. The solid lines in Fig. \ref{Fig4}(d)-(f) represent fits to this GL expression. From the resistivity measurements, the extracted zero-temperature upper critical fields $\mu_{0} H_{c2}(0)$ are 5.38(6), 2.84(2) and 2.17(4) T, with corresponding fitted $T_{c}$ values of 4.74(2), 3.64(1) and 3.01(1) K for Re$_{0.76}$V$_{0.24}$, Re$_{0.56}$Nb$_{0.44}$ and Re$_{0.60}$Ta$_{0.40}$ compounds, respectively. At finite fields, vortex motion and flux flow may broaden the resistive transition, making the 50 \% criterion less reliable for $H_{c2}$; magnetization and heat-capacity measurements may therefore provide more reliable estimates. Analysis of the magnetization data yields $\mu_{0}H_{c2}(0)$ values of 6.09(7) and 2.71(9) T at $T_{c}$ = 4.47(1) and 2.34(1), for Re$_{0.76}$V$_{0.24}$ and Re$_{0.56}$Nb$_{0.44}$ compounds. From the heat capacity measurements, the corresponding values of $\mu_{0} H_{c2}(0)$ are 6.57(7), 2.50(5) and 1.92(2) T and $T_{c}$ = 4.52(1), 4.44(2), and 1.76(1) K for the Re$_{0.76}$V$_{0.24}$, Re$_{0.56}$Nb$_{0.44}$ and Re$_{0.60}$Ta$_{0.40}$ compounds. For subsequent analysis of superconducting parameters, the $\mu_{0} H_{c2}(0)$ values obtained from the heat capacity measurements were used. These values are summarized in Table \ref{tbl:parameters}. 

The Pauli paramagnetic limiting field $\mu_{0} H^{P}_{c2}(0)$, which arises due to Zeeman splitting of Cooper pairs, is approximated as $\mu_{0}$H$_{p}$ = $\Delta_{0}/\sqrt{2} \mu_{B}$, where $\Delta_{0}$ is the superconducting energy gap at zero temperature and $\mu_{B}$ is the Bohr magneton \cite{chandrasekhar1962note, clogston1962upper}. Within the weak-coupling BCS framework, this expression can be written as $\mu_{0}$H$_{p}$ = 1.86$*$T$_{c}$, providing a convenient estimate of the Pauli limiting field directly from the $T_c$ values. Using value of $T_c$ the corresponding Pauli limiting fields were estimated to be 8.4, 4.5, and 3.3 T for Re$_{0.76}$V$_{0.24}$, Re$_{0.56}$Nb$_{0.44}$ and Re$_{0.60}$Ta$_{0.40}$ respectively. 

\begin{figure*}
\includegraphics[width=2.0\columnwidth]{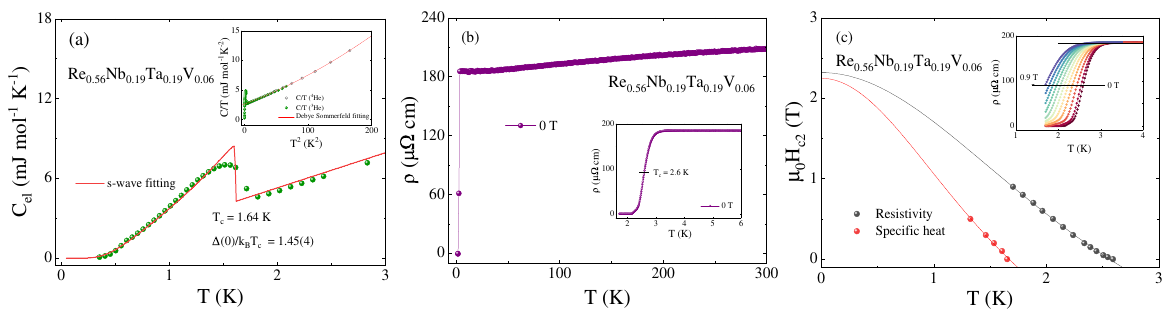}
\caption {\label{Fig6} (a) Temperature dependence of the electronic heat capacity C$_{el}$ as a function of T, extracted from total heat capacity and fitted with an isotropic s-wave superconducting gap model. The inset shows the zero-field heat capacity of Re$_{0.56}$Nb$_{0.19}$Ta$_{0.19}$V$_{0.06}$ plotted as C/T versus T$^2$, revealing a clear superconducting transition. (b) Temperature-dependent resistivity of Re$_{0.56}$Nb$_{0.19}$Ta$_{0.19}$V$_{0.06}$, with the inset highlighting a sharp superconducting transition. (c) Upper critical field $H_{c2}$, estimated from resistivity and heat capacity measurements, is shown and fitted using phenomenological Ginzburg–Landau (GL) interpolation framework. The inset presents temperature-dependent resistivity under applied magnetic fields, demonstrating the suppression of superconductivity.}
\end{figure*}

The Ginzburg-Landau (GL) coherence length, $\xi_{GL}$, was determined from upper critical field using the relation $H_{c2}(0) = \frac{\Phi_{0}}{2\pi\xi_{GL}^{2}}$, where $\Phi_{0}$ = $hc/2e$ = $2.07 \times 10^{-15}$ Wb is the flux quantum \cite{tinkham2004introduction}. By substituting the upper critical field values obtained from the heat capacity measurements into the above relation, the Ginzburg-Landau coherence lengths $\xi_{\mathrm{GL}}$ were estimated to be 71 \AA~, 114 \AA~ and 131 \AA~ for Re$_{0.76}$V$_{0.24}$, Re$_{0.56}$Nb$_{0.44}$ and Re$_{0.60}$Ta$_{0.40}$, respectively. The superconducting penetration depth $\lambda_{GL}(0)$ was subsequently evaluated by using lower critical field $H_{c1}(0)$ as well as the $\xi_{GL}$ values, through the given relation \cite{klimczuk2007physical, tinkham2004introduction}: 
\begin{equation}
H_{c1}(0) = \frac{\Phi_{0}}{4\pi\lambda_{GL}^2(0)}\left[ln \frac{\lambda_{GL}(0)}{\xi_{GL}(0)}\right].
\end{equation}
The resulting Ginzburg-Landau penetration depth $\lambda_{GL}$(0) were determined to be 3550 \AA~ for Re$_{0.76}$V$_{0.24}$ and 2980 \AA~ for Re$_{0.56}$Nb$_{0.44}$. Using the relation $\kappa_{GL}$ = $\lambda_{GL}(0)$/$\xi_{GL}(0)$ the corresponding Ginzburg-Landau parameters were found to be 50 and 26 for Re$_{0.76}$V$_{0.24}$ and Re$_{0.56}$Nb$_{0.44}$ compounds. Since $\kappa_{GL}$ $>$ 1/$\sqrt{2}$ A all cases, these values clearly confirm that the studied alloys belong to the class of type-II superconductors. Finally, the thermodynamic critical field $H_{c}$ was estimated using the relation $H_{c}^{2}\ln \kappa_{GL} = H_{c1}H_{c2}$. The obtained values are 92 mT for Re$_{0.76}$V$_{0.24}$ and 68 mT for Re$_{0.56}$Nb$_{0.44}$. Overall, the derived superconducting parameters such as $\lambda_{GL}$, $\xi_{GL}$, $H_{c}$, and $\kappa_{GL}$ are summarized in Table \ref{tbl:parameters}. Some superconducting parameters for Re$_{0.60}$Ta$_{0.40}$ could not be determined because of its low superconducting transition temperature and the consequent lack of magnetization data.

\subsection{Medium entropy alloy of Re-Nb-Ta-V}
Motivated by the growing interest in medium- and high-entropy alloy systems as platforms for superconductivity in strongly disordered material, we synthesized and investigated the superconducting properties of the Re–Nb–Ta–V medium-entropy alloy (MEA). The large configurational entropy in these alloys promotes chemical disorder while stabilizing solid-solution phases. Here, we report a Re-based MEA derived from the structurally complex $\sigma$ phases of Re-X, which is stable only within a narrow compositional range. The optimized composition, Re$_{0.56}$Nb$_{0.19}$Ta$_{0.19}$V$_{0.06}$, yields a configurational entropy of mixing of 1.124$R$, significantly higher than in conventional alloys. Achieving a single-phase material was particularly challenging because of the limited stability range of the parent $\sigma$ phase. A single-phase formation was obtained only at a V concentration of 6 at. \% V. Even slight deviations from this optimized composition, i.e., further increases in the V content, resulted in the formation of secondary phases, indicating that the single-phase medium-entropy alloy exists only within a very narrow compositional window. The powder X-ray diffraction (XRD) patterns of the synthesized samples are shown in Fig. \ref{Fig5}(a), confirming the formation of a single-phase tetragonal $\sigma$-phase crystal structure. The lattice parameters obtained from Le Bail refinement are $a = 9.7309(3)$ \AA~and $c = 5.0650(7)$ \AA~with V$_{cell}$ = 479.60(6) \AA$^{3}$ and are larger than those refined for Re$_{0.76}$V$_{0.24}$ but smaller than those reported for Re$_{0.60}$Ta$_{0.40}$. For this MEA, meaningful Rietveld refinement is limited by substantial site mixing/disorder, which prevents reliable refinement of individual site occupancies. The EDS result confirms that MEA is close to the nominal compositions, while scanning electron microscopy (SEM) reveals a uniform microstructure. The corresponding results are shown in \ref{Fig5}(b).

Further, the superconducting properties of Re$_{0.56}$Nb$_{0.19}$Ta$_{0.19}$V$_{0.06}$ were investigated using electrical resistivity and heat capacity measurements. A clear jump in the heat capacity at the superconducting transition temperature confirms bulk superconductivity with $T_c = 1.64$ K, as shown in the inset of Fig. \ref{Fig6}(a). The normal-state heat capacity data were analyzed using the procedure described in the previous section, yielding a Sommerfeld coefficient $\gamma_n = 2.59(7)$ mJ mol$^{-1}$ K$^{-2}$, a Debye coefficient $\beta_3 = 0.040(2)$ mJ mol$^{-1}$ K$^{-4}$, and an anharmonic contribution $\beta_5$ = 0.087(4) $\mu$J mol$^{-1}$ K$^{-6}$. From these parameters, the Debye temperature was estimated to be $\Theta_D = 365$ K. The electron-phonon coupling constant was evaluated as $\lambda_{\mathrm{el-ph}} = 0.47$, indicating weak electron-phonon coupling in this medium-entropy alloy. Having both $\gamma_n$ and $\lambda_{\mathrm{el-ph}}$ the electronic density of states at the Fermin energy was found to be $DOS(E_F) = 0.75$ st. eV$^{-1}$ f.u.$^{-1}$. The electronic heat capacity in the superconducting state was analyzed using a single-gap $s$-wave model, as described for the binary compounds, and is shown in Fig. \ref{Fig6}(a). The fit yields a superconducting gap ratio $\alpha = \Delta(0)/k_B T_c = 1.45(4)$, which is smaller than the weak-coupling BCS value of 1.74, and may suggest deviations from conventional single-gap behavior. Using the Sommerfeld coefficient, the normalized specific-heat jump was determined to be $\Delta C/\gamma_n T_c = 1.05(2)$ for Re$_{0.56}$Nb$_{0.19}$Ta$_{0.19}$V$_{0.06}$, lower than the BCS weak-coupling value.

Electrical transport measurements were performed to further confirm the superconducting nature of the Re$_{0.56}$Nb$_{0.19}$Ta$_{0.19}$V$_{0.06}$ alloy. The electrical resistivity data exhibit a transition to zero resistance at $T_c$, providing clear evidence of superconductivity, as shown in Fig. \ref{Fig6}(b). The determined $T_c$ using the 50 \% resistivity drop criterion is 2.59 K. Similar to Re$_{0.56}$Nb$_{0.44}$ and Re$_{0.6}$Ta$_{0.4}$ alloys, superconducting critical temperature revealed from the resistivity is higher than that determined from the heat capacity measurement. Notably, the low-temperature resistivity upturn is strongly suppressed in Re$_{0.56}$Nb$_{0.44}$ and completely suppressed in Re$_{0.56}$Nb$_{0.19}$Ta$_{0.19}$V$_{0.06}$, possibly due to the increased compositional complexity and chemical/structural disorder. Temperature-dependent resistivity measurements under various applied magnetic fields are presented in inset of the Fig. \ref{Fig6}(c). Analysis of these data using the procedure described in the previous section yields an upper critical field of $\mu_{0}H_{c2}(0) = 2.31(2)$ T, with a corresponding $T_c$ = 2.56(3) K, obtained from resistivity. Since resistivity measurements may be influenced by surface superconductivity effects, the upper critical field was also estimated from specific-heat measurements, which more directly probe the bulk superconducting state. From heat capacity method, $\mu_{0}H_{c2}(0) = 2.24(3)$ T was obtained at $T_c = 1.67(2)$ K for the medium-entropy alloy. Accordingly, the $\mu_{0}H_{c2}(0)$ value derived from specific-heat data was used for the evaluation of the remaining superconducting parameters. The temperature dependence of $\mu_{0} H_{c2}$ obtained from both resistivity and specific-heat measurements is shown in Fig. \ref{Fig6}(c), and all extracted superconducting parameters are summarized in Table \ref{tbl:parameters}.

\section{Conclusion}
In summary, polycrystalline samples of Re–X (X = V, Nb, Ta) intermetallic compounds with compositions Re$_{0.76}$V$_{0.24}$, Re$_{0.56}$Nb$_{0.44}$, and Re$_{0.60}$Ta$_{0.40}$, along with a derived medium-entropy alloy (MEA) Re$_{0.56}$Nb$_{0.19}$Ta$_{0.19}$V$_{0.06}$, were synthesized via arc melting. Powder X-ray diffraction confirms that all samples crystallize in a single-phase tetragonal $\sigma$-phase structure. Magnetization, electrical resistivity, and specific-heat measurements establish bulk type-II superconductivity in all compounds, with transition temperatures $T_c$ = 4.46, 2.28, 1.76 and 1.64 K for Re$_{0.76}$V$_{0.24}$, Re$_{0.56}$Nb$_{0.44}$, Re$_{0.60}$Ta$_{0.40}$, and Re$_{0.56}$Nb$_{0.19}$Ta$_{0.19}$V$_{0.06}$, respectively. Analysis of thermodynamic and magnetic data indicates weakly coupled, conventional type-II superconductivity within the phenomenological Ginzburg-Landau interpolation framework. A systematic suppression of $T_c$ is observed upon changing transition metal X from 3d to 5d, as well as in the disordered MEA, which may be associated with change of the spin–orbit coupling strength, variations in valence electron concentration, and chemical disorder. The decrease in Debye temperature for heavier transition metals may also suppress $T_c$ by reducing the characteristic phonon energy scale, although this effect is coupled to variations in the electronic density of states and electron-phonon coupling strength. Recently, medium- and high-entropy Re- and Re–Os-based alloys have attracted considerable attention \cite{Huixia_1, Huixia_2}. These alloys adopt a non-centrosymmetric $\alpha$-Mn-type crystal structure, and their superconducting critical temperature has been found to vary linearly with the valence electron count (VEC). Among the compounds investigated here, Re$_{0.76}$V$_{0.24}$ has the highest VEC (6.52 el.), and its critical temperature is comparable to those reported in refs. \cite{Huixia_1, Huixia_2} for compounds with similar VEC values. Re$_{0.60}$Ta$_{0.40}$ and Re$_{0.56}$Nb$_{0.44}$, Re$_{0.56}$Nb$_{0.19}$Ta$_{0.19}$V$_{0.06}$ exhibit lower VEC = 6.12 and 6.20 el., and correspondingly lower critical temperatures. 

Overall, this work provides a systematic study of superconductivity in Re-based binary and derived medium-entropy $\sigma$-phase alloys, highlighting the interplay between electronic structure, chemical disorder, and superconductivity in structurally complex intermetallic systems. Despite strong spin–orbit coupling associated with Re and inherent structural disorder, the superconducting state remains consistent with conventional weak-coupling behavior. Further microscopic investigations, such as muon spin rotation and relaxation ($\mu$SR), would be valuable to clarify the superconducting gap structure and pairing symmetry. The stabilization of the $\sigma$-phase medium-entropy alloy further suggests a pathway toward predictive compositional tuning and data-driven design of medium- and high-entropy $\sigma$-phase materials, enabling systematic exploration of disorder effects on superconductivity and guiding future materials design.

\section{Acknowledgments} 
This project was supported by the National Science Center (Poland), Project No. 2022/45/B/ST5/03916. The authors would like to acknowledge Oliwia Barra for her contribution and support during the experiment. 

\section{Data Availability} 
The data are available from the authors upon reasonable request.

\bibliography{Ref}

@article{hall1966sigma,
  title={The sigma phase},
  author={Hall, EO and Algie, SH},
  journal={Metallurgical reviews},
  volume={11},
  number={1},
  pages={61--88},
  year={1966},
  publisher={Taylor \& Francis},
  url = {https://www.tandfonline.com/doi/abs/10.1179/mtlr.1966.11.1.61}
}

@article{xu2022superconductivity,
  title={Superconductivity and phase diagrams of CaK (Fe 1- x Mn x) 4 As 4 single crystals},
  author={Xu, M and Schmidt, J and Gati, E and Xiang, L and Meier, WR and Kogan, Vladimir G and Bud'ko, Sergey L and Canfield, PC},
  journal={Physical Review B},
  volume={105},
  number={21},
  pages={214526},
  year={2022},
  publisher={APS},
  url = {https://journals.aps.org/prb/abstract/10.1103/PhysRevB.105.214526}
}

@article{shoemaker1950crystal,
  title={The crystal structure of a sigma phase, \text{FeCr$^1$}},
  author={Shoemaker, David P and Bergman, Bror Gunnar},
  journal={Journal of the American Chemical Society},
  volume={72},
  number={12},
  pages={5793--5793},
  year={1950},
  publisher={ACS Publications},
  url = {https://pubs.acs.org/doi/pdf/10.1021/ja01168a550}
}

@article{bergman1954determination,
  title={The determination of the crystal structure of the $\sigma$ phase in the iron--chromium and iron--molybdenum systems},
  author={Bergman, Gunnar and Shoemaker, David P},
  journal={Acta Crystallographica},
  volume={7},
  number={12},
  pages={857--865},
  year={1954},
  publisher={International Union of Crystallography},
  url = {https://journals.iucr.org/paper?s0365110x54002605}
}

@article{yaqoob2012comparison,
  title={Comparison of the site occupancies determined by combined Rietveld refinement and density functional theory calculations: Example of the ternary \text{Mo--Ni--Re} $\sigma$ phase},
  author={Yaqoob, Khurram and Crivello, Jean-Claude and Joubert, Jean-Marc},
  journal={Inorganic Chemistry},
  volume={51},
  number={5},
  pages={3071--3078},
  year={2012},
  publisher={ACS Publications},
  url = {https://pubs.acs.org/doi/full/10.1021/ic202479y}
}

@article{mandal2025time,
  title={Time-reversal symmetry breaking in a \text{Re}-based kagome lattice superconductor},
  author={Mandal, Manasi and Kataria, A and Meena, PK and Kushwaha, RK and Singh, D and Biswas, PK and Stewart, R and Hillier, AD and Singh, RP},
  journal={Physical Review B},
  volume={111},
  number={5},
  pages={054511},
  year={2025},
  publisher={APS},
  url = {https://journals.aps.org/prb/abstract/10.1103/PhysRevB.111.054511}
}

@article{singh2014detection,
  title={Detection of time-reversal symmetry breaking in the noncentrosymmetric superconductor \text{Re$_6$Zr} using muon-spin spectroscopy},
  author={Singh, Ravi P and Hillier, Adrian D and Mazidian, B and Quintanilla, J and Annett, JF and Paul, D McK and Balakrishnan, Geetha and Lees, Martin R},
  journal={Physical review letters},
  volume={112},
  number={10},
  pages={107002},
  year={2014},
  publisher={APS},
  url = {https://journals.aps.org/prl/abstract/10.1103/PhysRevLett.112.107002}
}

@article{kasper1956ordering,
  title={Ordering of atoms in the $\sigma$ phase},
  author={Kasper, JS and Waterstrat, RM},
  journal={Acta Crystallographica},
  volume={9},
  number={3},
  pages={289--295},
  year={1956},
  publisher={International Union of Crystallography},
  url = {https://journals.iucr.org/paper?S0365110X56000802}
}

@article{compton1961superconductivity,
  title={Superconductivity of technetium alloys and compounds},
  author={Compton, VB and Corenzwit, E and Maita, JP and Matthias, BT and Morin, FJ},
  journal={Physical Review},
  volume={123},
  number={5},
  pages={1567},
  year={1961},
  publisher={APS},
  url = {https://journals.aps.org/pr/abstract/10.1103/PhysRev.123.1567}
}

@article{khan1977comparison,
  title={Comparison of superconducting and electronic properties of $\sigma$-and A-15 phase \text{Nb-Al}},
  author={Khan, HR and Raub, Ch J and L{\"u}ders, K and Sz{\"u}cs, Z},
  journal={Applied physics},
  volume={13},
  number={2},
  pages={123--129},
  year={1977},
  publisher={Springer},
  url = {https://link.springer.com/article/10.1007/BF00882469}
}

@article{khan1979superconductivity,
  title={Superconductivity of A15-and $\sigma$-phase of \text{Nb-Ir}},
  author={Khan, HR and Raub, Ch J and L{\"u}ders, K and Sz{\"u}cs, Z},
  journal={Applied physics},
  volume={19},
  number={2},
  pages={231--235},
  year={1979},
  publisher={Springer},
  url = {https://link.springer.com/article/10.1007/BF00932403}
}

@article{rasmussen1987new,
  title={New crystal data for \text{Mo$_{.63}$Ru$_{.37}$ (Mo$_5$Ru$_3$)} a superconducting sigma phase},
  author={Rasmussen, Svend Erik and Lundtoft, Britta},
  journal={Powder Diffraction},
  volume={2},
  number={1},
  pages={29--30},
  year={1987},
  publisher={Cambridge University Press},
  url = {https://www.cambridge.org/core/journals/powder-diffraction/article/abs/new-crystal-data-for-mo63ru37-mo5ru3-a-superconducting-sigma-phase/FD1258694148973D0CA7A9FF8C57E1A9}
}

@article{khan1978comparison,
  title={Comparison of superconducting parameters of \text{A15}-and $\sigma$-phases of \text{Nb- Pt}},
  author={Khan, HR and Raub, Ch J and L{\"u}ders, K and Sz{\"u}cs, Z},
  journal={Applied physics},
  volume={15},
  number={3},
  pages={307--313},
  year={1978},
  publisher={Springer}, 
  url = {https://link.springer.com/article/10.1007/BF00896113}
}

@article{singhal1968formation,
  title={The formation of ferrite and sigma-phase in some austenitic stainless steels},
  author={Singhal, LK and Martin, JW},
  journal={Acta Metallurgica},
  volume={16},
  number={12},
  pages={1441--1451},
  year={1968},
  publisher={Elsevier},
  url = {https://www.sciencedirect.com/science/article/abs/pii/0001616068900394}
}

@article{jorda1986vanadium,
  title={The vanadium-rhenium system: phase diagram and superconductivity},
  author={Jorda, JL and Muller, J},
  journal={Journal of the Less Common Metals},
  volume={119},
  number={2},
  pages={337--345},
  year={1986},
  publisher={Elsevier},
  url = {https://www.sciencedirect.com/science/article/abs/pii/0022508886906946}
}

@article{bucher1961supraleitung,
  title={Supraleitung und Paramagnetismus in komplexen Phasen der {\"U}bergangsmetalle},
  author={Bucher, E and Heiniger, F and M{\"u}ller, J},
  journal={Helv Phys Acta},
  volume={34},
  pages={843--858},
  year={1961}, 
  url = {https://link.springer.com/article/10.1007/BF02422846}
}

@article{joubert2008crystal,
  title={Crystal chemistry and Calphad modeling of the $\sigma$ phase},
  author={Joubert, J-M},
  journal={Progress in Materials Science},
  volume={53},
  number={3},
  pages={528--583},
  year={2008},
  publisher={Elsevier}, 
  url = {https://www.sciencedirect.com/science/article/abs/pii/S0079642507000242}
}

@article{roberts1976survey,
  title={Survey of superconductive materials and critical evaluation of selected properties},
  author={Roberts, Benjamin Washington},
  journal={Journal of Physical and Chemical Reference Data},
  volume={5},
  number={3},
  pages={581--822},
  year={1976},
  publisher={American Institute of Physics for the National Institute of Standards and~…}, 
  url = {https://pubs.aip.org/aip/jpr/article-abstract/5/3/581/242140/Survey-of-superconductive-materials-and-critical}
}

@article{knapton1959niobium,
  title={The niobium-rhenium system},
  author={Knapton, AG},
  journal={Journal of the Less Common Metals},
  volume={1},
  number={6},
  pages={480--486},
  year={1959},
  publisher={Elsevier},
  url = {https://www.sciencedirect.com/science/article/abs/pii/0022508859900657}
}

@article{hsieh2012overview,
  title={Overview of intermetallic sigma ($\sigma$) phase precipitation in stainless steels},
  author={Hsieh, Chih-Chun and Wu, Weite},
  journal={International Scholarly Research Notices},
  volume={2012},
  number={1},
  pages={732471},
  year={2012},
  publisher={Wiley Online Library},
  url = {https://onlinelibrary.wiley.com/doi/full/10.5402/2012/732471}
}

@article{tavares2010magnetic,
  title={Magnetic detection of sigma phase in duplex stainless steel \text{UNS S31803}},
  author={Tavares, SSM and Pardal, JM and Guerreiro, JL and Gomes, AM and Da Silva, MR},
  journal={Journal of Magnetism and Magnetic Materials},
  volume={322},
  number={17},
  pages={L29--L33},
  year={2010},
  publisher={Elsevier},
  url = {https://www.sciencedirect.com/science/article/pii/S0304885310001629}
}

@article{carnicom2017new,
  title={New $\sigma$-phases in the \text{Nb--X--Ga} and \text{Nb--X--Al} systems \text{(X= Ru, Rh, Pd, Ir, Pt, and Au)}},
  author={Carnicom, Elizabeth M and Xie, Weiwei and Klimczuk, Tomasz and Cava, Robert J},
  journal={Dalton Transactions},
  volume={46},
  number={41},
  pages={14158--14163},
  year={2017},
  publisher={Royal Society of Chemistry},
  url = {https://pubs.rsc.org/en/content/articlehtml/2017/dt/c7dt02955a}
}

@article{khan1980superconductivity,
  title={Superconductivity and resistance behaviour of $\sigma$-phase alloys: \text{Nb-Rh} and \text{Ta-Rh}},
  author={Khan, HR and L{\"u}ders, K and Raub, Ch J and Roth, G},
  journal={Zeitschrift f{\"u}r Physik B Condensed Matter},
  volume={38},
  number={1},
  pages={27--33},
  year={1980},
  publisher={Springer},
  url = {https://link.springer.com/article/10.1007/BF01321199}
}

@article{fisk1973normal,
  title={Normal state resistance behavior and superconductivity},
  author={Fisk, Z and Lawson, AC},
  journal={Solid State Communications},
  volume={13},
  number={3},
  pages={277--279},
  year={1973},
  url = {https://escholarship.org/content/qt0pc3s2j0/qt0pc3s2j0.pdf}
}

@article{hulm1972superconductivity,
  title={Superconductivity in the \text{TiO and NbO} systems},
  author={Hulm, JK and Jones, CK and Hein, RA and Gibson, JW},
  journal={Journal of Low Temperature Physics},
  volume={7},
  number={3},
  pages={291--307},
  year={1972},
  publisher={Springer},
  url = {https://link.springer.com/article/10.1007/bf00660068}
}

@article{hake1961electrical,
  title={Electrical resistivity, Hall effect and superconductivity of some bcc titanium-molybdenum alloys $*$},
  author={Hake, RR and Leslie, DH and Berlincourt, TG},
  journal={Journal of Physics and Chemistry of Solids},
  volume={20},
  number={3-4},
  pages={177--186},
  year={1961},
  publisher={Elsevier},
  url = {https://www.sciencedirect.com/science/article/abs/pii/0022369761900026}
}

@article{chandrasekhar1961paramagnetic,
  title={The paramagnetic susceptibilities of some uranium-molybdenum alloys between 100° K and 300° K},
  author={Chandrasekhar, BS and Bardeen, JM},
  journal={Journal of Physics and Chemistry of Solids},
  volume={21},
  number={3-4},
  pages={206--209},
  year={1961},
  publisher={Elsevier}, 
  url = {https://www.sciencedirect.com/science/article/abs/pii/0022369761900993}
}

@article{cohen1967effect,
  title={Effect of Fermi-level motion on normal-state properties of $\beta$-tungsten superconductors},
  author={Cohen, Roger W and Cody, GD and Halloran, John J},
  journal={Physical Review Letters},
  volume={19},
  number={15},
  pages={840},
  year={1967},
  publisher={APS}, 
  url = {https://journals.aps.org/prl/abstract/10.1103/PhysRevLett.19.840}
}

@misc{roberts1967intermetallic,
  title={Intermetallic Compounds},
  author={Roberts, BW},
  year={1967},
  publisher={Wiley, New York}
}

@article{carnicom2017superconductivity,
  title={Superconductivity in the \text{Nb-Ru-Ge} $\sigma$ phase},
  author={Carnicom, Elizabeth M and Xie, Weiwei and Sobczak, Zuzanna and Kong, Tai and Klimczuk, Tomasz and Cava, Robert J},
  journal={Physical Review Materials},
  volume={1},
  number={7},
  pages={074802},
  year={2017},
  publisher={APS},
  url = {https://journals.aps.org/prmaterials/abstract/10.1103/PhysRevMaterials.1.074802}
}

@article{blaugher1961superconductivity,
  title={Superconductivity in the $\sigma$ and $\alpha$-\text{Mn} structures},
  author={Blaugher, RD and Hulm, JK},
  journal={Journal of Physics and Chemistry of Solids},
  volume={19},
  number={1-2},
  pages={134--138},
  year={1961},
  publisher={Elsevier},
  url = {https://www.sciencedirect.com/science/article/abs/pii/0022369761900671}
}

@article{koch1971superconductivity,
  title={Superconductivity in \text{Mo-Re} and \text{Nb-Ir} $\sigma$ phases},
  author={Koch, CC and Scarbrough, JO},
  journal={Physical Review B},
  volume={3},
  number={3},
  pages={742},
  year={1971},
  publisher={APS},
  url = {https://journals.aps.org/prb/abstract/10.1103/PhysRevB.3.742}
}

@article{klimczuk2004carbon,
  title={Carbon isotope effect in superconducting \text{MgCNi$_3$}},
  author={Klimczuk, Tomasz and Cava, Robert Joseph},
  journal={Physical Review B—Condensed Matter and Materials Physics},
  volume={70},
  number={21},
  pages={212514},
  year={2004},
  publisher={APS},
  url = {https://journals.aps.org/prb/abstract/10.1103/PhysRevB.70.212514}
}

@article{carnicom2018sigma,
  title={The $\sigma$-phase superconductors \text{Nb$_{20.4}$Rh$_{5.7}$Ge$_{3.9}$} and \text{Nb$_{20.4}$Rh$_{5.7}$Si$_{3.9}$}},
  author={Carnicom, Elizabeth M and Kong, Tai and Klimczuk, Tomasz and Cava, Robert J},
  journal={Solid State Communications},
  volume={284},
  pages={96--101},
  year={2018},
  publisher={Elsevier},
  url = {https://www.sciencedirect.com/science/article/pii/S0038109818302709}
}

@article{mcmillan1968transition,
  title={Transition temperature of strong-coupled superconductors},
  author={McMillan, W. L.},
  journal={Phys. Rev.},
  volume={167},
  number={2},
  pages={331},
  year={1968},
  publisher={APS},
  url={https://journals.aps.org/pr/abstract/10.1103/PhysRev.167.331}
}

@article{padamsee1973quasiparticle,
  title={Quasiparticle phenomenology for thermodynamics of strong-coupling superconductors},
  author={Padamsee, H. and Neighbor, J. E. and Shiffman, C. A.},
  journal={J. Low Temp. Phys.},
  volume={12},
  pages={387},
  year={1973},
  publisher={Springer},
url={https://link.springer.com/article/10.1007/BF00654872}
}

@article{chandrasekhar1962note,
  title={A note on the maximum critical field of high-field superconductors},
  author={Chandrasekhar, B. S.},
  journal={Appl. Phys. Letters},
  volume={1},
  year={1962},
  publisher={Westinghouse Research Labs., Pittsburgh},
  url={https://doi.org/10.1063/1.1777362}
}

@article{clogston1962upper,
  title={Upper limit for the critical field in hard superconductors},
  author={Clogston, A. M.},
  journal={Phys. Rev. Lett.},
  volume={9},
  number={6},
  pages={266},
  year={1962},
  publisher={APS},
  url={https://journals.aps.org/prl/abstract/10.1103/PhysRevLett.9.266}
}

@book{tinkham2004introduction,
  title={Introduction to superconductivity},
  author={Tinkham, M.},
  volume={1},
  year={2004},
  publisher={Courier Corporation}
}

@article{ghosh2021recent,
  title={Recent progress on superconductors with time-reversal symmetry breaking},
  author={Ghosh, Sudeep Kumar and Smidman, Michael and Shang, Tian and Annett, James F and Hillier, Adrian D and Quintanilla, Jorge and Yuan, Huiqiu},
  journal={Journal of Physics: Condensed Matter},
  volume={33},
  number={3},
  pages={033001},
  year={2021},
  publisher={IOP Publishing},
  url = {https://iopscience.iop.org/article/10.1088/1361-648X/abaa06/meta}
}

@article{shang2018time,
  title={Time-reversal symmetry breaking in \text{Re}-based superconductors},
  author={Shang, Tian and Smidman, Michael and Ghosh, Saikat K and Baines, Christopher and Chang, Lieh-Jeng and Gawryluk, DJ and Barker, Joel AT and Singh, Ravi P and Paul, D McK and Balakrishnan, Geetha and others},
  journal={Physical review letters},
  volume={121},
  number={25},
  pages={257002},
  year={2018},
  publisher={APS},
  url = {https://journals.aps.org/prl/abstract/10.1103/PhysRevLett.121.257002}
}

@article{shang2019structure,
  title={Structure and superconductivity in the binary \text{Re$_{1-x}$Mo$_x$} alloys},
  author={Shang, Tian and Gawryluk, Dariusz J and Verezhak, Joel AT and Pomjakushina, Ekaterina and Shi, Ming and Medarde, Marisa and Mesot, Jo{\"e}l and Shiroka, Toni},
  journal={Physical Review Materials},
  volume={3},
  number={2},
  pages={024801},
  year={2019},
  publisher={APS},
  url = {https://journals.aps.org/prmaterials/abstract/10.1103/PhysRevMaterials.3.024801}
}

@article{shang2020re1,
  title={\text{Re$_{1-x}$Mo$_{x}$} as an ideal test case of time-reversal symmetry breaking in unconventional superconductors},
  author={Shang, Tian and Baines, Christopher and Chang, Lieh-Jeng and Gawryluk, Dariusz Jakub and Pomjakushina, Ekaterina and Shi, Ming and Medarde, Marisa and Shiroka, Toni},
  journal={npj Quantum Materials},
  volume={5},
  number={1},
  pages={76},
  year={2020},
  publisher={Nature Publishing Group UK London},
  url = {https://www.nature.com/articles/s41535-020-00279-1}
}

@article{poole1995superconductivity,
  title={Superconductivity Academic Press},
  author={Poole Jr, Charles P and Farach, HA and Creswick, RJ and Prozorov, R},
  journal={New York},
  volume={154},
  year={1995},
  url = {https://scholar.google.com/scholar?hl=en&as_sdt=0%2C5&q=Poole%2C+C.+P.+Jr.%2C+Farach%2C+H.+A.%2C+Creswick%2C+R.+J.+%26+Prozorov%2C+R.+Superconductivity+%28Academic+Press%2C+2010%29.&btnG=}
}

@article{bardeen1957theory,
  title={Theory of superconductivity},
  author={Bardeen, John and Cooper, Leon N and Schrieffer, John Robert},
  journal={Physical review},
  volume={108},
  number={5},
  pages={1175},
  year={1957},
  publisher={APS},
  url ={https://journals.aps.org/pr/abstract/10.1103/PhysRev.108.1175}
}

@article{meena2026superconducting,
  title={Superconducting Properties of \text{Re$_2$Zr$_{0.5}$Hf$_{0.5}$} and \text{Re$_6$Zr$_{0.5}$Hf$_{0.5}$} Alloys},
  author={Meena, Pavan Kumar and Arushi and Singh, Ravi Prakash},
  journal={physica status solidi (a)},
  volume={223},
  number={1},
  pages={e202500721},
  year={2026},
  publisher={Wiley Online Library},
  url = {https://onlinelibrary.wiley.com/doi/full/10.1002/pssa.202500721}
}

@article{kushwaha2024unconventional,
  title={Unconventional properties of the noncentrosymmetric superconductor \text{Re$_{8}$NbTa}},
  author={Kushwaha, RK and Arushi and Sharma, S and Srivastava, S and Meena, PK and Pula, M and Beare, J and Gautreau, J and Hillier, AD and Luke, GM and others},
  journal={Physical Review B},
  volume={109},
  number={17},
  pages={174518},
  year={2024},
  publisher={APS},
  url = {https://journals.aps.org/prb/abstract/10.1103/PhysRevB.109.174518}
}

@article{klimczuk2007physical,
  title={Physical properties of the noncentrosymmetric superconductor \text{$Mg_{10}Ir_{19}B_{16}$}},
  author={Klimczuk, Tomasz and Ronning, Filip and Sidorov, Vladimir and Cava, Robert J and Thompson, Joe David},
  journal={Physical review letters},
  volume={99},
  number={25},
  pages={257004},
  year={2007},
  publisher={APS},
  url = {https://journals.aps.org/prl/abstract/10.1103/PhysRevLett.99.257004}
}

@article{han2013superconductivity,
  title={Superconductivity and strong intrinsic defects in \text{LaPd$_{1-x}$Bi$_{2}$}},
  author={Han, Fei and Malliakas, Christos D and Stoumpos, Constantinos C and Sturza, Mihai and Claus, Helmut and Chung, Duck Young and Kanatzidis, Mercouri G},
  journal={Physical Review B—Condensed Matter and Materials Physics},
  volume={88},
  number={14},
  pages={144511},
  year={2013},
  publisher={APS}, 
  url = {https://journals.aps.org/prb/abstract/10.1103/PhysRevB.88.144511}
}

@article{kushwaha2026microscopic,
  title={Microscopic study of superconductivity in the noncentrosymmetric $\alpha$-Mn alloy \text{NbTaOs$_2$}},
  author={Kushwaha, RK and Arushi and Jangid, S and Meena, PK and Stewart, R and Hillier, AD and Singh, RP},
  journal={Physical Review B},
  volume={113},
  number={9},
  pages={094510},
  year={2026},
  publisher={APS}, 
  url = {https://journals.aps.org/prb/abstract/10.1103/hpj7-qnzq}
}

@article{tsindlekht2004tunneling,
  title={Tunneling and magnetic characteristics of superconducting \text{ZrB$_{12}$} single crystals},
  author={Tsindlekht, MI and Leviev, GI and Asulin, I and Sharoni, A and Millo, O and Felner, I and Paderno, Yu B and Filippov, VB and Belogolovskii, MA},
  journal={Physical Review B—Condensed Matter and Materials Physics},
  volume={69},
  number={21},
  pages={212508},
  year={2004},
  publisher={APS}, 
  url = {https://journals.aps.org/prb/abstract/10.1103/PhysRevB.69.212508}
}

@article{Huixia_1,
  title={Strongly correlated electronic superconductivity in the noncentrosymmetric Re-Os-based high/medium-entropy alloys},
  author={Chen, Rui and Li, Longfu and Zeng, Lingyong and Li, Kuan and Yu, Peifeng and Wang, Kangwang and Xiang, Zaichen and Wang, Shuangyue and Qin, Jingjun and Zhang, Wanyi and others},
  journal={Acta Materialia},
  pages={121401},
  year={2025},
  publisher={Elsevier},
  url = {https://www.sciencedirect.com/science/article/pii/S1359645425006871}
}

@article{Huixia_2,
  title={Superconductivity in the medium-entropy/high-entropy Re-based alloys with a non-centrosymmetric $\alpha$-Mn lattice},
  author={Li, Kuan and Li, Longfu and Zeng, Lingyong and Li, Yucheng and Chen, Rui and Yu, Peifeng and Wang, Kangwang and Xiang, Zaichen and Shang, Tian and Luo, Huixia},
  journal={Superconductor Science and Technology},
  volume={38},
  number={5},
  pages={055010},
  year={2025},
  publisher={IoP Publishing},
  url = {https://iopscience.iop.org/article/10.1088/1361-6668/adc8df/meta}
}

\end{document}